# Twelve-spin "Schrödinger cat"

Jae-Seung Lee and A. K. Khitrin

Department of Chemistry, Kent State University, Kent OH 44242-0001

**Abstract**

Pseudopure "cat" state, a superposition of quantum states with all spins up and all spins down, is experimentally demonstrated for a system of twelve dipolar-coupled nuclear spins of fully $^{13}$C-labeled benzene molecule oriented in a liquid-crystalline matrix.

Quantum entanglement, a form of non-classical correlation, has been an important issue in recent debates on the foundations of quantum mechanics.[1] In quantum information science, the entanglement is considered as a physical resource, which plays a crucial role in quantum teleportation,[2] quantum key distribution,[3,4] and quantum computing.[5,6]

A cat state is a special case of entanglement: superposition of the two most distinct states (alive and dead). Building a cat state is a benchmark for controlling quantum systems.[7] The decoherence time of the cat state sets limitation on quantum computing because this state is supposed to be the most fragile among quantum states. Cat states are the central elements in high-precision spectroscopy,[8] amplified quantum detection[9,10] and measurement.[11] In such applications, possible gain from using the cat states is proportional to the number of particles which can be entangled. In this letter, we describe an experiment on creating a pseudopure cat state of a system of twelve dipolar-coupled nuclear spins. At present, it is the largest and the most complex composite quantum system where manipulations with individual states have been performed.

The physical system is a 5% solution of fully labeled $^{13}C_6$-benzene (Aldrich) in liquid-crystalline solvent MLC-6815 (EMD Chemical). Each benzene molecule contains twelve nuclear spins, six $^{13}C$ and six protons, coupled by residual dipole-dipole interactions. All intermolecular spin-spin interactions are averaged out by fast molecular motions. Therefore, the system is an ensemble of non-interacting spin clusters, where each benzene molecule contains twelve dipolar-coupled nuclear spins. The experiment has been performed with a Varian Unity/Inova 500 MHz NMR spectrometer.

For further discussion it will be convenient to use compact notations for spin states. The symbols $|u\rangle_H = |\uparrow\uparrow\uparrow\uparrow\uparrow\uparrow\rangle_H$ and $|d\rangle_H = |\downarrow\downarrow\downarrow\downarrow\downarrow\downarrow\rangle_H$ will denote the states of the proton spins with all six spins up and down, respectively. Similarly, the notations $|u\rangle_C$ and $|d\rangle_C$ will be used for the $^{13}C$ spin states with all six carbon spins up or down.

The experimental scheme is shown in Fig. 1. It consists of the three parts: preparation of the pseudopure state (step A), creation of the twelve-spin cat state (steps B-E), verification and life-time estimation (steps F-J). The pulse sequence in step A initializes our system in a pseudopure state $|\Psi_A\rangle = |d\rangle_C |d\rangle_H$. Preparation of the pseudopure state includes multiple-quantum evolution, filtering with a double array of phases and evolution times, and partial saturation. This technique is described in detail elsewhere.[12] The unit of the evolution time $\tau = 65.6$ μs corresponds to a rotation, caused by the hetero-nuclear dipolar couplings, of the protons' 6Q coherence by $\pi$ when the $^{13}C$ spins are in the state $|u\rangle_C$ or $|d\rangle_C$. The $^1H$ linear-response spectrum of $|\Psi_A\rangle$ is presented in Fig. 2(a). Due to the high symmetry of the benzene molecule, the spectrum of $|\Psi_A\rangle$ has only one $^1H$ peak and one $^{13}C$ peak (not shown). Thermal equilibrium $^1H$ and $^{13}C$ spectra of this system have thousands of unresolved peaks. Starting from $|\Psi_A\rangle$, one can obtain other pseudopure states by using hard and selective pulses. As an example, 180° hard pulses on both protons and carbons convert the state $|\Psi_A\rangle$ into the pseudopure state $|u\rangle_C |u\rangle_H$ (Fig. 2(b)).

To create the twelve-spin cat state, we expanded the scheme used to build the seven-spin cat state in benzene with a single $^{13}C$ label.[11] For the seven-spin system, a single carbon spin played a role of a control register to change the state of proton spins. In this work, the control register consists of six carbon spins, and some modification of the pulse sequence is necessary. In step B, instead of a single pulse, 10 cycles of a 16-pulse sequence were applied to convert the state $|d\rangle_C$ into a superposition of $|u\rangle_C$ and $|d\rangle_C$. The 16-pulse sequence,[13] with slightly better performance for carbons, evolves the system with the same double-quantum Hamiltonian as the 8-pulse sequence[14] used for the proton spins. Therefore, with optimized parameters, the 16-pulse sequence transforms $|\Psi_A\rangle$ into $|\Psi_B\rangle = (|u\rangle_C + |d\rangle_C)|d\rangle_H/\sqrt{2}$. After the 8-pulse sequence in step C brings the proton spins into a superposition, the state becomes $|\Psi_C\rangle = (|u\rangle_C + |d\rangle_C)(|u\rangle_H + |d\rangle_H)/2$. Free evolution in step D after the time $\tau/2$ gives phase factors according to a sum of the total z-components of proton and carbon spins: $|\Psi_D\rangle = (|u\rangle_C|u\rangle_H - i|u\rangle_C|d\rangle_H - i|d\rangle_C|u\rangle_H + |d\rangle_C|d\rangle_H)/2$. Finally, the 8-pulse sequence in step E converts $|\Psi_D\rangle$ into the cat state $|\Psi_E\rangle = (|u\rangle_C|u\rangle_H + |d\rangle_C|d\rangle_H)/\sqrt{2}$. The spectrum for this state, shown in Fig. 2(c), is consistent with the equally-weighted superposition of $|u\rangle_C|u\rangle_H$ and $|d\rangle_C|d\rangle_H$. However, the linear-response spectrum, produced by the small-angle reading pulse, depends only on the two diagonal elements of the corresponding density matrix. The other two, off-diagonal, elements of the cat state density matrix constitute the twelve-quantum coherence which is not directly observable.

Therefore, we added the state-verification steps by time-reversing the sequences of the steps B-E. The pulse sequences and their phases were arranged so that steps G, H, I, and J are respectively the inverses of the steps E, D, C, and B. The delay step F was added for measuring decoherence time of the cat state. With no delay in step F, we expected that the state would be converted back to the state $|\Psi_A\rangle$. Fig. 2(d) shows the result of such experiment. It confirms that the state $|\Psi_E\rangle$ after the step E is a superposition of the states $|u\rangle_C|u\rangle_H$ and $|d\rangle_C|d\rangle_H$ rather than their mixture. The peak marked by an asterisk in Figs. 2(c) and 2(d) comes from an impurity, which gives very high sharp peak in the thermal equilibrium spectrum.

It was difficult to measure accurately the cat state decoherence time. There are several reasons for that. The pulse sequence is very long: there are 1301 radio-frequency and gradient pulses in the total sequence. Therefore, despite considerable efforts to tune its parameters, accumulation of errors was unavoidable. Measurements were performed near the sensitivity limit and instabilities of the spectrometer during long acquisition times affected the accuracy. Due to weaker carbon-carbon interactions, the multiple-quantum carbon subsequences were considerably longer than the ones for protons (8.8 ms vs. 3.2 ms, respectively). As a result, quality of decoupling from protons during the carbon subsequences was insufficient and caused significant loss of the signal.

In addition to the verification part, consisting of steps G-K in Fig.1, we tried several other schemes including phase-incremented sequences with subsequent Fourier transform for measuring the decay of intensity of the twelve-quantum coherence in the state $|\Psi_E\rangle$. All our measurements gave unexpectedly short lifetime of 3.2 to 4.7 ms. We are not absolutely sure that it is a true lifetime of the cat state but, at the same time, we cannot suggest any explanations of how any of the non-idealities mentioned above could cause such a short decay time.

For comparison, we present below some accurately measured relaxation times for this and similar systems. The decay time for the diagonal elements of the cat state, measured with varying delays in step F but without the time-reversed sequences of steps G-J, was found to be 0.30 s. Conventionally measured $T_1$ and $T_2$ relaxation times are 1.7 s and 0.25 s for $^1$H and 2.5 s and 0.26 s for $^{13}$C, respectively. The decay time of protons' 6Q coherence was measured when the 6Q coherence was excited starting with one of the two different initial conditions: the thermal equilibrium state and the pseudopure state $|\Psi_A\rangle$. For the thermal equilibrium initial state, we averaged the decay times for different peaks. The decay times were 70 ms and 16 ms for the thermal equilibrium and pseudopure initial states, respectively. This difference is a general feature of relaxation in composite systems: when (pseudo)pure state relaxes, the intensity only leaves the state; for relaxation of the same state in a mixture, intensities of coherences or populations of levels are transferred in both directions.

In similar systems of natural-abundance benzene (six spins) and single-labeled benzene (seven spins) in liquid crystalline matrix, the relaxation times are the following. For a six-spin system of natural-abundance benzene, average $T_1$, $T_{1\rho}$, $T_2$, and the decay time of the pseudopure ground state (all spins up) and 6Q coherence are 3.30 s, 1.75 s, 0.47 s, 0.75 s, and 0.087 s, respectively. For a seven-spin system of single-labeled benzene, $T_1$ and $T_2$ of individual peaks are 1.7 – 2.3 s and 0.1 – 0.7 s for proton spins and 1.4 – 2.9 s and 0.3 – 1.1 s for carbon spins, respectively. The decay time of the protons' 6Q coherence, excited from the thermal equilibrium state, relaxation times of the diagonal and off-diagonal elements of the seven-spin cat state are 0.073 s, 0.49 s, and 0.029 s, respectively.

In summary, a twelve-spin cat state has been experimentally demonstrated in a system of dipolar-coupled nuclear spins of fully $^{13}$C-labeled benzene in liquid crystalline matrix. In terms of the number of qubits, it is the largest "Schrödinger cat" ever built.

**Figure Captions**

Fig. 1. Experimental pulse sequence. (Step A) preparation of the pseudopure state; (steps B-E) creation of the twelve-spin cat state; (steps F-J) state verification and life-time estimation.

Fig. 2. $^1$H Spectra: (a) pseudopure state |d>|d>; (b) pseudopure state |u>|u> obtained from |d>|d> by applying 180° hard pulses to both proton and carbon spins; (c) a cat state, superposition of |d>|d> and |u>|u>; (d) |d>|d> after the time-reversed sequences of entangling operations.

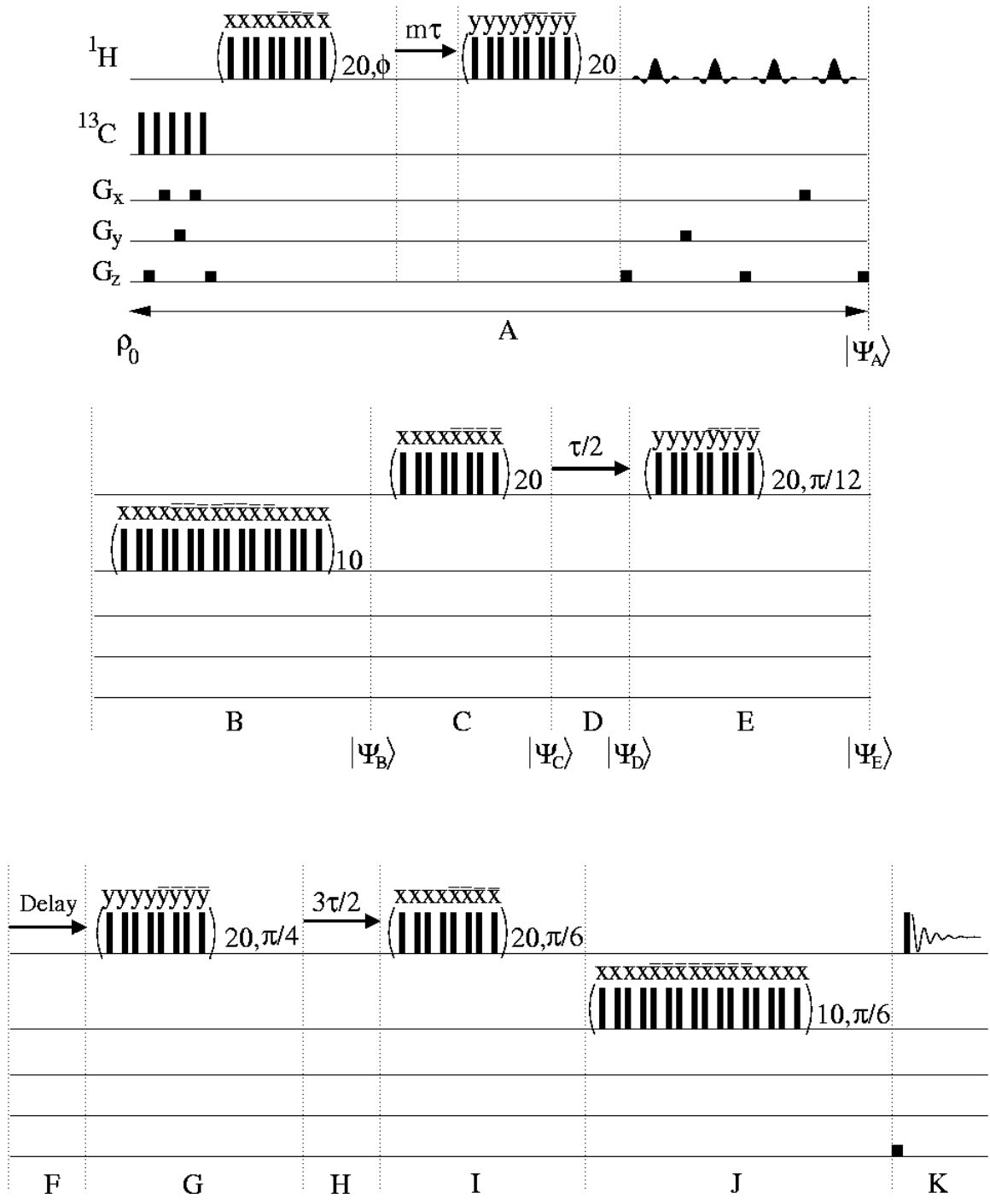

Fig. 1

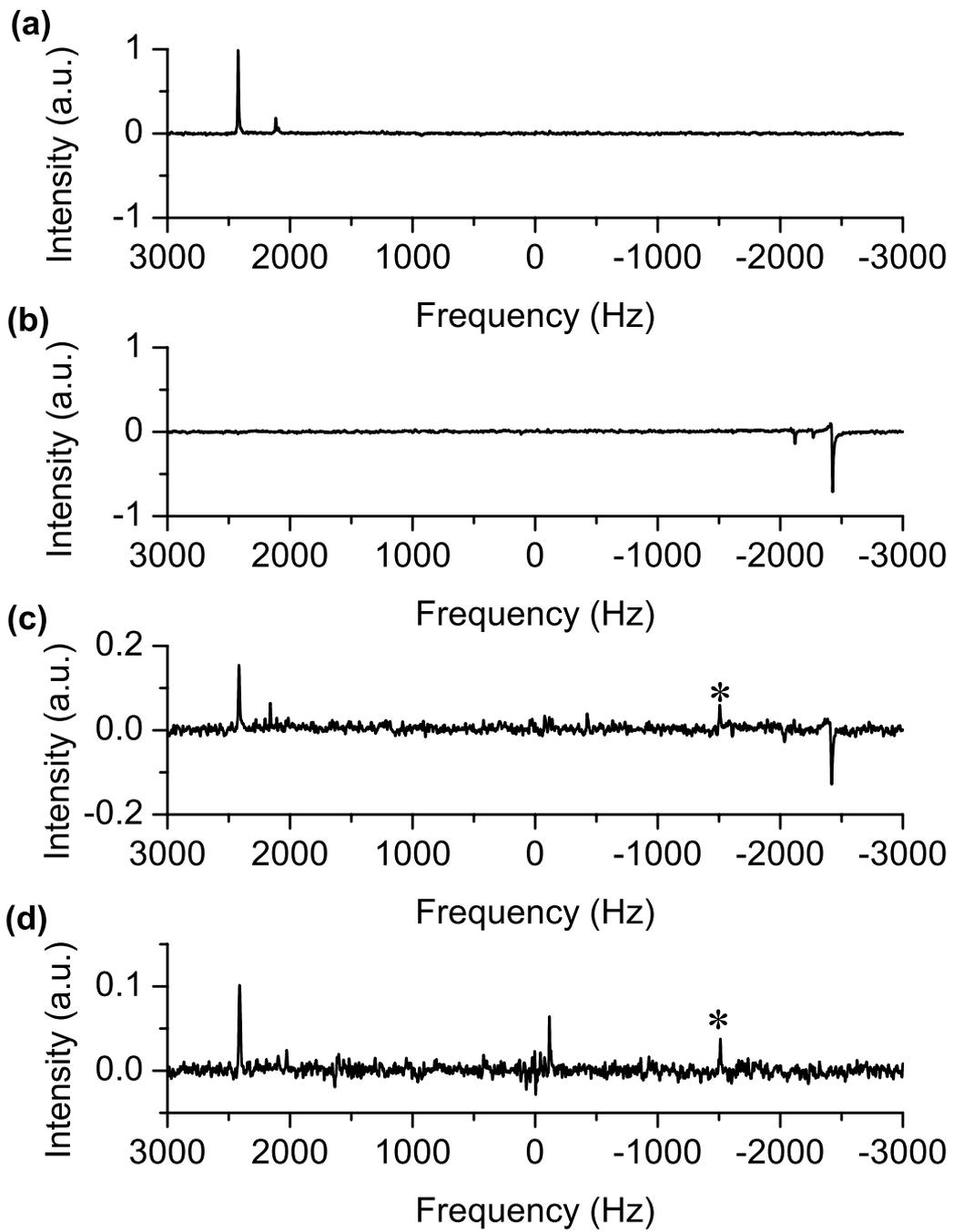

Fig. 2